\documentclass[%
    reprint,
    superscriptaddress,
    amsmath,
    amssymb,
    aps
]{revtex4-2}

\usepackage{graphicx}
\usepackage{dcolumn}
\usepackage{bm}
\usepackage{siunitx}
\usepackage[dvipsnames]{xcolor}
\usepackage[colorlinks]{hyperref}
\usepackage{placeins}
\hypersetup{%
    plainpages=true,
    breaklinks=true,
    hypertexnames=false,
    pageanchor=true,
    colorlinks=true,
    linkcolor={green},
    citecolor={blue},
    urlcolor={blue},
    anchorcolor={black}
}

\begin{document}

\preprint{APS/123-QED}

\title{Evolution of $1/f$ Flux Noise in Superconducting Qubits with Weak Magnetic Fields}

\def\affilPhysics{Department of Physics, Massachusetts Institute of Technology, Cambridge, MA 02139, USA}
\def\affilLL{MIT Lincoln Laboratory, Lexington, MA 02421, USA}
\def\affilRLE{Research Laboratory of Electronics, Massachusetts Institute of Technology, Cambridge, MA 02139, USA}
\def\affilPhysicsHarvard{Department of Physics, Harvard University, Cambridge, MA 02139, USA}
\def\affilAQ{\textit{Atlantic Quantum, Cambridge, MA 02139}}
\def\affilEECS{Department of Electrical Engineering and Computer Science, Massachusetts Institute of Technology, Cambridge, MA 02139, USA}

\author{David~A.~Rower}
  \email{rower@mit.edu}
  \affiliation{\affilPhysics}
  \affiliation{\affilRLE}
  
\author{Lamia~Ateshian}%
  \affiliation{\affilRLE}
  \affiliation{\affilEECS}
  
\author{Lauren~H.~Li}%
  \affiliation{\affilPhysics}
  
\author{Max~Hays}%
  \affiliation{\affilRLE}
  
\author{Dolev~Bluvstein}%
  \affiliation{\affilPhysicsHarvard}
  
\author{Leon~Ding}%
  \affiliation{\affilPhysics}
  \affiliation{\affilRLE}
  
\author{Bharath~Kannan}%
    \altaffiliation[Present address: ]{\affilAQ}
\author{Aziza~Almanakly}%
  \affiliation{\affilRLE}
  \affiliation{\affilEECS}
  
\author{Jochen~Braum\"uller}%
  \affiliation{\affilRLE}
  
\author{David~K.~Kim}%
\author{Alexander~Melville}%
\author{Bethany~M.~Niedzielski}%
\author{Mollie~E.~Schwartz}%
\author{Jonilyn~L.~Yoder}%
  \affiliation{\affilLL}
  
\author{Terry~P.~Orlando}%
  \affiliation{\affilRLE}
  \affiliation{\affilEECS}
  
\author{Joel~I-Jan~Wang}%
\author{Simon Gustavsson}%
  \altaffiliation[Additional address: ]{\affilAQ}
\author{Jeffrey~A.~Grover}%
  \affiliation{\affilRLE}

\author{Kyle~Serniak}%
  \affiliation{\affilRLE}
  \affiliation{\affilLL}
  
\author{Riccardo~Comin}%
  \affiliation{\affilPhysics}
  
\author{William~D.~Oliver}%
  \affiliation{\affilPhysics}
  \affiliation{\affilRLE}
  \affiliation{\affilEECS}
  \affiliation{\affilLL}
 
\date{\today}

\begin{abstract}
The microscopic origin of $1/f$ magnetic flux noise in superconducting circuits has remained an open question for several decades despite extensive experimental and theoretical investigation. Recent progress in superconducting devices for quantum information has highlighted the need to mitigate sources of qubit decoherence, driving a renewed interest in understanding the underlying noise mechanism(s). Though a consensus has emerged attributing flux noise to surface spins, their identity and interaction mechanisms remain unclear, prompting further study. Here we apply weak in-plane magnetic fields to a capacitively-shunted flux qubit (where the Zeeman splitting of surface spins lies below the device temperature) and study the flux-noise-limited qubit dephasing, revealing previously unexplored trends that may shed light on the dynamics behind the emergent $1/f$ noise. Notably, we observe an enhancement (suppression) of the spin-echo (Ramsey) pure dephasing time in fields up to $B=\SI{100}{G}$. With direct noise spectroscopy, we further observe a transition from a $1/f$ to approximately Lorentzian frequency dependence below \SI{10}{Hz} and a reduction of the noise above \SI{1}{MHz} with increasing magnetic field. We suggest that these trends are qualitatively consistent with an increase of spin cluster sizes with magnetic field. These results should help to inform a complete microscopic theory of $1/f$ flux noise in superconducting circuits.
\end{abstract}

\maketitle

The experimental progress towards building quantum processors with superconducting qubits has advanced significantly in recent years. 
However, environmental noise and material quality limit qubit coherence, which constrains the ability to scale to larger devices and utilize different qubit architectures~\cite{kjaergaardSuperconductingQubitsCurrent2020, gambettaBuildingLogicalQubits2017, siddiqiEngineeringHighcoherenceSuperconducting2021}. 
One major limitation to qubit coherence is the ubiquitous low-frequency magnetic-flux noise which displays a $1/f$ power spectral density~\cite{wellstoodLowFrequencyNoise1987, paladinoNoiseImplicationsSolidstate2014}. 
This noise often limits the dephasing time of frequency-tunable qubits~\cite{yoshiharaDecoherenceFluxQubits2006, bialczakFluxNoiseJosephson2007, kakuyanagiDephasingSuperconductingFlux2007, kochChargeInsensitiveQubit2007, bylanderNoiseSpectroscopyDynamical2011} and the fidelity of flux-activated gates~\cite{mccourtLearningNoiseDynamical2022}. 
Removing the source of $1/f$ flux noise would greatly expand the design space for next-generation quantum hardware, yet the origin of the noise has remained an open question for decades.

Several microscopic theories of magnetic defects in superconducting circuits with emergent $1/f$ flux noise spectra have been proposed~\cite{paladinoNoiseImplicationsSolidstate2014, mullerUnderstandingTwolevelsystemsAmorphous2019, degraafChemicalStructuralIdentification2022, navaaquinoFluxNoiseDisordered2022}.
However, there is a lack of consensus in the community on both the nature and source of the spins and the spin physics which gives rise to the noise.
Nonetheless, several experimental constraints for microscopic flux noise models have been established, including an emergent $1/f^\alpha$ noise power spectral density from $10^{-4}$ \SI{}{Hz} to $10^{8}$ \SI{}{Hz} with $\alpha \lesssim 1$~\cite{ quintanaObservationClassicalQuantumCrossover2017, yanFluxQubitRevisited2016}, anticorrelation of the noise in loops sharing a boundary~\cite{gustavssonNoiseCorrelationsFlux2011}, perimeter scaling of the noise amplitude~\cite{lantingGeometricalDependenceLowfrequency2009, braumullerCharacterizingOptimizingQubit2020}, pivoting of the noise spectrum with temperature about a fixed frequency~\cite{antonMagneticFluxNoise2013}, non-vanishing flux-inductance noise cross-correlation~\cite{sendelbachComplexInductanceExcess2009}, paramagnetic temperature dependence of the spin bath susceptibility~\cite{sendelbachMagnetismSQUIDsMillikelvin2008}, and asymmetry of the noise spectrum~\cite{quintanaObservationClassicalQuantumCrossover2017}.
Several of these features point to the likely relevance of spin-spin interactions~\cite{sendelbachMagnetismSQUIDsMillikelvin2008, degraafDirectIdentificationDilute2017} and emergent phenomena including spin diffusion~\cite{lantingEvidenceTemperaturedependentSpin2014, lantingProbingEnvironmentalSpin2020} and clustering~\cite{antonMagneticFluxNoise2013, sendelbachComplexInductanceExcess2009,  lantingProbingEnvironmentalSpin2020}.
In addition to providing low-frequency dephasing noise, magnetic defects may also play a role in broadband flux noise which contributes to high-frequency energy relaxation processes~\cite{bylanderNoiseSpectroscopyDynamical2011, yanFluxQubitRevisited2016, quintanaObservationClassicalQuantumCrossover2017}, or give rise to other decoherence mechanisms~\cite{degraafSuppressionLowfrequencyCharge2018, graafTwolevelSystemsSuperconducting2020}.

Despite the extensive experimental and theoretical efforts to understand and mitigate $1/f$ flux noise, one critical characterization has remained absent: the response of the flux-noise spectrum to magnetic fields.
Such characterization proves experimentally challenging due to the interplay of magnetic fields with superconductors comprising  devices (often Al or Nb metallizations on Si or sapphire substrates) and the isolation of flux noise from other noise sources~\cite{wangMeasurementControlQuasiparticle2014, schneiderTransmonQubitMagnetic2019, luthiEvolutionNanowireTransmon2018, winkelImplementationTransmonQubit2020, kringhojMagneticFieldCompatibleSuperconductingTransmon2021, krauseMagneticFieldResilience2022, samkharadzeHighKineticInductanceSuperconductingNanowire2016, degraafFrequencyNoiseSuperconducting2018, borisovSuperconductingGranularAluminum2020, graafTwolevelSystemsSuperconducting2020}.

In this Letter, we investigate $1/f$ flux noise as a function of applied magnetic fields up to $B=\SI{100}{G}$ with a superconducting flux qubit, where the field is oriented in the plane of the device. At low frequencies ($\lesssim\SI{10}{Hz}$), we observe a $1/f$ to approximately Lorentzian transition in the noise spectrum accompanied by an increase of the Ramsey pure-dephasing rate with applied field. Surprisingly, at high frequencies ($\gtrsim\SI{1}{MHz}$) we observe a suppression of the flux noise and an increase in the $1/f^\alpha$ noise exponent $\alpha$ with applied field. These results provide the first study to date of flux-noise-limited qubit dephasing and $1/f$ flux noise evolution in magnetic fields, which can serve as a new experimental reference for future microscopic theories of flux noise.

\begin{figure}[t!]
    \centering
    \includegraphics[width=0.9\columnwidth]{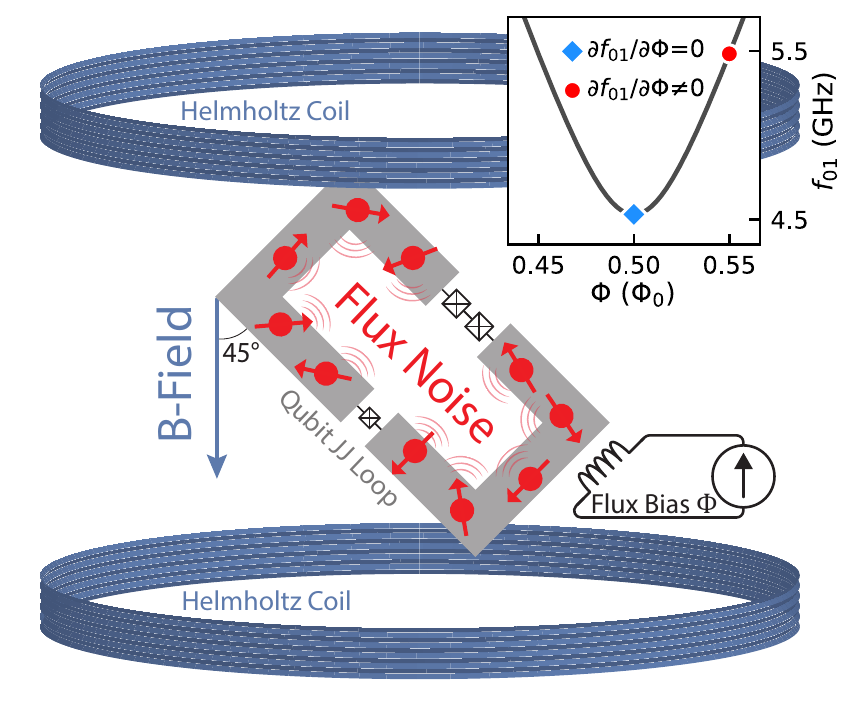}
    
    \caption{\textbf{Flux qubit in a magnetic field.} A simplified schematic of the experimental setup. The Josephson-junction (JJ) loop (gray) of a flux qubit is placed in a uniform magnetic field provided by Helmholtz coils (blue). The field is oriented in the plane of the device, and the device is tilted at a $45^\circ$ in-plane angle relative to the field. Surface spins (red) in proximity to the loop generate flux noise which dephases the qubit. The inset shows an example flux qubit spectrum, with frequency $f_{01}$ as a function of an independent flux bias $\Phi$. The blue diamond indicates the point of first-order flux insensitivity (the so-called ``sweet spot''). The red circle highlights an example operating point where the qubit displays flux-noise-limited dephasing.}
    \label{fig:fig1}
\end{figure} 

\begin{figure*}[t!]
    \centering
    \includegraphics[width=\textwidth]{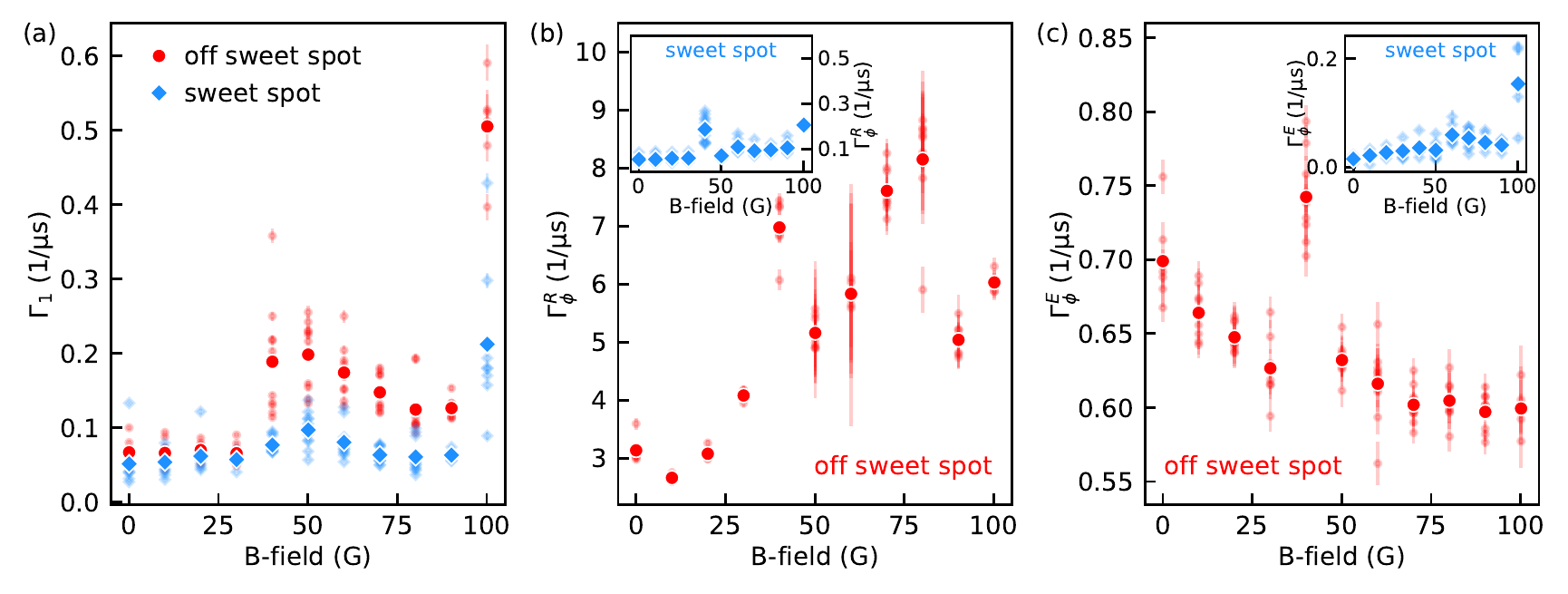}
    
    \caption{\textbf{Evolution of qubit coherence with an in-plane magnetic field.} Data taken at the sweet spot ($\partial f_{01}/\partial \Phi = 0$, blue diamonds) and off the sweet spot ($|\partial f_{01}/\partial \Phi| = \SI{26.0}{\giga Hz / \Phi_0}$, red circles). \textbf{a)} Energy relaxation rate $\Gamma_1$. \textbf{b)} Ramsey pure-dephasing rate $\Gamma^{R}_{\phi}$. \textbf{c)} Spin-echo pure dephasing rate $\Gamma^{E}_{\phi}$. Insets in \textbf{b)}, \textbf{c)} show dephasing rates at the sweet spot. Data was taken during nine field sweeps, with $\Gamma_1$, $\Gamma^{R}_{\phi}$, and $\Gamma^{E}_{\phi}$ measured consecutively at each bias point and field. Individual rate measurements are presented as partially transparent small markers with error bars given by the fit uncertainty. Average rates at each field are presented as large opaque markers. The outlier dephasing at $B=\SI{40}{G}$ is likely dominated by noise in the applied field (see Supplemental Material~\cite{supp} for details).}
    \label{fig:fig2}
\end{figure*}

We measured capacitively-shunted flux qubit samples comprising Al metalizations with Al/AlOx/Al Josephson junctions (JJs) on a Si substrate at the base temperature of a dilution refrigerator with $T\lesssim\SI{40}{\milli K}$. The samples were mounted on a cold finger with superconducting magnets in a Helmholtz coil geometry providing an approximately in-plane magnetic field (where we estimate the out-of-plane component to be $\approx0.2\%$ of the total applied field). The sample was rotated $45^\circ$ relative to the field direction in order to ensure the qubit was sensitive to spin fluctuations both along and transverse to the direction of the field. The experimental setup and a representative qubit spectrum are shown in Fig.~\ref{fig:fig1} (see Supplemental Material for details~\cite{supp}). 

To probe broad flux-noise trends with magnetic field, we first performed standard qubit coherence measurements of the energy-relaxation rate $\Gamma_1 = 1/T_1$ and pure-dephasing rates from Ramsey ($\Gamma_{\phi}^R$) and spin-echo ($\Gamma_{\phi}^E$) protocols. We characterized the qubit coherence both at the flux degeneracy point (where the qubit is first-order insensitive to flux noise, i.e. $ \partial f_{01} /  \partial \Phi = 0$, hereafter referred to as the ``sweet spot''), and at a flux bias where dephasing was dominated by flux noise ($| \partial f_{01} /  \partial \Phi| = \SI{26.0}{\giga Hz/\Phi_0}$, calibrated at each field with an independent flux control) [Fig.~\ref{fig:fig2}]. In order to extract the pure dephasing trends, we first isolated the field dependence of $\Gamma_1$ [Fig.~\ref{fig:fig2}a].
We found that $\Gamma_1$ varies non-monotonically, but generally increases with field. These observations may be due to the softening of the Al superconducting gap at higher fields and an associated elevated population of quasiparticles (QPs), or the effects of vortices penetrating the thin-film aluminum of our device~\cite{wangMeasurementControlQuasiparticle2014, luthiEvolutionNanowireTransmon2018, nsanzinezaTrappingSingleVortex2014}. We also observed a slight difference in $\Gamma_1$ at the two different working points, which is accounted for in the analysis of the qubit pure dephasing.

We extracted the Ramsey and spin-echo pure dephasing rates as proxies for the low- and high-frequency flux-noise power. 
Off the sweet spot, the Ramsey and spin-echo decay envelopes were approximately Gaussian and therefore consistent with $1/f$-limited dephasing~\cite{braumullerCharacterizingOptimizingQubit2020}. We fit the decays to the product of an exponential envelope from energy relaxation and a Gaussian envelope from pure dephasing, with the relaxation rate fixed from an immediately preceding measurement (see Supplemental Material for details~\cite{supp}). With increasing field, we observed an increase in the quasistatic noise power probed by $\Gamma^R_\phi$ [Fig.~\ref{fig:fig2}b] accompanied by a decrease in the $\gtrsim \SI{}{MHz}$ noise probed by $\Gamma^E_\phi$ [Fig.~\ref{fig:fig2}c]. At the sweet spot, Ramsey and spin-echo traces followed exponential decays and were therefore not $1/f$ limited. We observed relaxation-limited spin-echo dephasing ($\Gamma_\phi^{E} \lesssim \Gamma_1 / 2$) and Ramsey dephasing of the same order as the relaxation rate ($\Gamma_\phi^{R} \sim \Gamma_1$).  All coherence data were taken in nine separate runs, during each of which the field was first swept from $B=\SI{0}{G}$ to $B=\SI{100}{G}$ and then reversed. No hysteretic behavior was observed in $\Gamma^{R/E}_\phi$.

\begin{figure*}[t!]
    \centering
    \includegraphics[width=\textwidth]{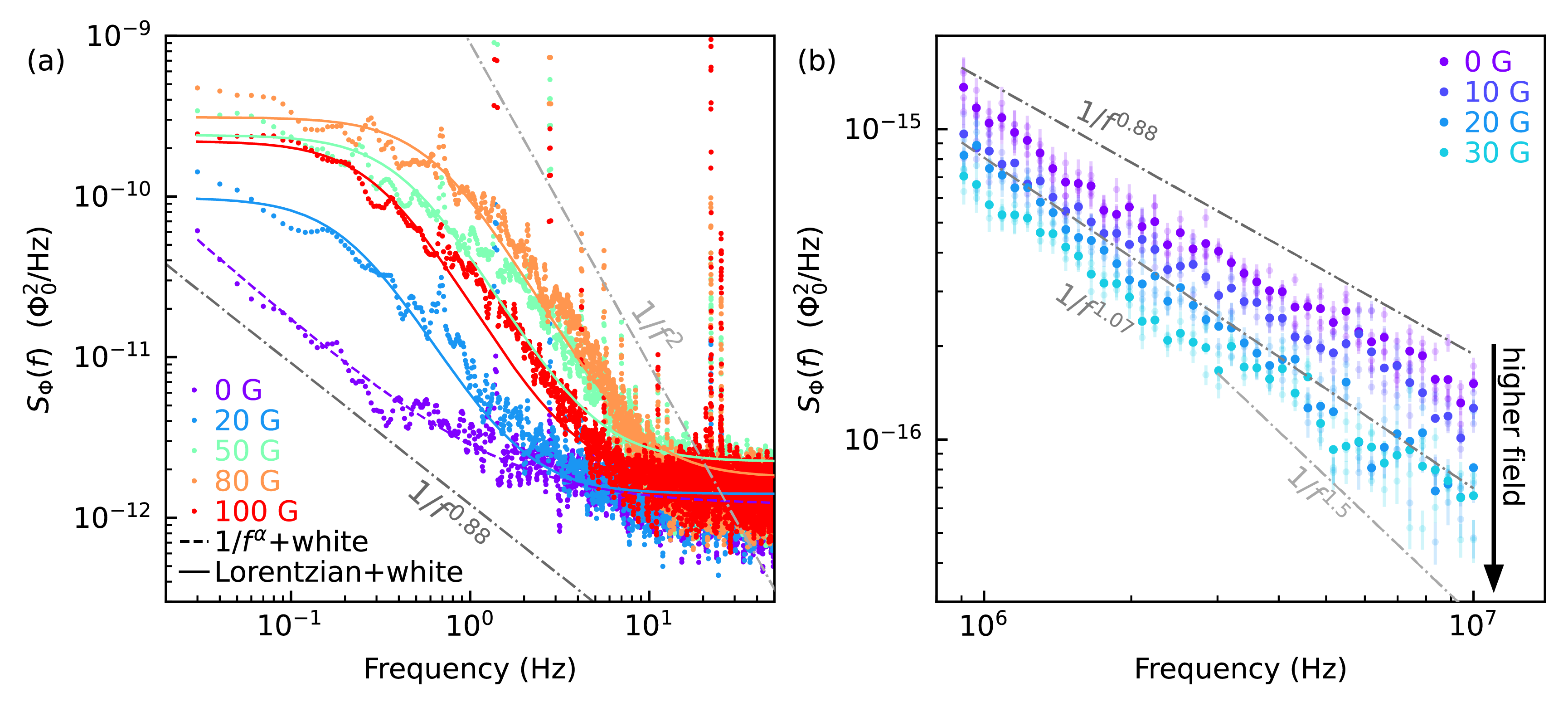}
    
    \caption{\textbf{Evolution of flux noise with an in-plane magnetic field.}  \textbf{a)} Low-frequency noise spectroscopy taken with single-shot Ramsey measurements. Data for $B<\SI{100}{G}$ were taken in one upwards sweep with $|\partial f_{01} / \partial \Phi| = \SI{22.0}{\giga Hz / \Phi_0}$, and data at $B=\SI{100}{G}$ was taken in a separate upwards sweep with $|\partial f_{01} / \partial \Phi| = \SI{21.0}{\giga Hz / \Phi_0}$. Gray dash-dotted lines serve as guides to the eye displaying  power laws $1/f^{0.88}$ (bottom) and $1/f^{2}$ (top, characteristic of a Lorentzian roll-off). The $B=\SI{0}{G}$ data is fit to a $1/f$ + white noise model (purple, dashed line), and data at each non-zero field is fit to a Lorentzian + white noise model (solid line, color of corresponding data). We attribute the white noise floor to readout infidelity. For further details, see Supplemental Material~\cite{supp}. \textbf{b)} Spin-locking noise spectroscopy. Data was taken in four separate field sweeps with $|\partial f_{01} / \partial \Phi| = \SI{30.0}{\giga Hz/\Phi_0}$ for $B\lesssim\SI{10}{G}$ and $|\partial f_{01} / \partial \Phi| = \SI{31.0}{\giga Hz/\Phi_0}$ for $B\gtrsim\SI{20}{G}$. Individual measurements are presented as partially transparent small markers with error bars given by the spin-locking decay fit uncertainty. Averages at each field are presented with opaque markers. Gray dash-dotted lines serve as guides to the eye displaying the power laws $1/f^\alpha$ with $\alpha=0.88$ (top, $\alpha$ from fit to $B=\SI{0}{G}$ data), $\alpha=1.07$ (middle, from fit to $B=\SI{20}{G}$ data), and $1/f^{1.5}$ (bottom, characteristic of the asymptotic behavior of spin-diffusion noise). At higher fields, we note a suppression of the measured flux noise, denoted by an annotated black arrow.}
    \label{fig:fig3}
\end{figure*}

To gain further insight into the nature of the flux noise trends, we measured the noise power spectral density (PSD) as a function of magnetic field. For low frequencies ($\lesssim\SI{10}{Hz}$), we utilized the single-shot Ramsey technique described in~\cite{yanSpectroscopyLowfrequencyNoise2012}, and for high frequencies ($\gtrsim\SI{1}{MHz}$), we utilized the spin-locking technique detailed in~\cite{yanRotatingframeRelaxationNoise2013} (see Supplemental Material~\cite{supp} for additional details). We observed an increase in the low-frequency noise along with a $1/f$ ($B=\SI{0}{G}$) to approximately Lorentzian ($B\gtrsim\SI{20}{G}$) transition in the PSD [Fig.~\ref{fig:fig3}a]. We emphasize that, in contrast to the general noise increase, the noise appears to decrease from $B=\SI{80}{G}$ to $B=\SI{100}{G}$, which is also present in the $\Gamma_\phi^{R}$ trend. Surprisingly, we also observed beating in Ramsey decays at intermediate fields $\SI{50}{G} \lesssim B \lesssim \SI{90}{G}$ (shown in Supplemental Material Fig.~S4~\cite{supp}) which may be consistent with telegraphic noise processes giving rise to the corresponding Lorentzian-like spectra; we leave a confirmation of the consistency between these observations to follow-up studies. At high frequencies, we observed a suppression of the flux noise in fields up to $B=\SI{30}{G}$ [Fig.~\ref{fig:fig3}b] (past this field, high-fidelity calibration for spin-locking spectroscopy became difficult due to the excess low-frequency noise). Both low- and high-frequency PSD trends were reproduced with a second qubit on the same chip.
To confirm that flux noise was responsible for the observed trends, we measured qubit frequency noise at the sweet spot and found it primarily magnetic-field-independent and well below the off-sweet-spot noise in the frequency ranges of interest [Fig.~S5a,c]. 
We also observed slight hysteretic behavior of the flux noise PSD at low frequencies [Fig.~S5b] but not at high frequencies [Fig.~S5d].
We note that both noise spectroscopy methods measure the symmetrized PSD of qubit frequency fluctuations, $S_{f_{01}}(f)$, and when operating away from the sweet spot, we utilized the conversion between frequency- and flux-noise PSDs $S_{f_{01}}(f) = (\partial f_{01} / \partial \Phi)^2 \cdot S_\Phi(f)$. To validate this conversion, we confirmed the echo dephasing rate varied linearly with the flux noise susceptibility $\partial f_{01} / \partial \Phi$ (as in~\cite{braumullerCharacterizingOptimizingQubit2020}) at multiple magnetic fields. 

We now discuss possible physical mechanisms that could explain our observations. We first explore the relevance of spin polarization with the applied field, which depends on temperature and is expected to reduce total SQUID flux noise power~\cite{laforestFluxvectorModelSpin2015}. Similar experiments have observed evidence for the low-frequency ($h f \ll k_B T_\text{eff}$) environment of the spin bath being in thermal equilibrium at an effective temperature $T_{\text{eff}}$ close to but above that of the mixing-chamber plate~\cite{quintanaObservationClassicalQuantumCrossover2017}. Studies of the native surface spin bath of Al$_2$O$_3$ observed signatures consistent with a population of $g=2$, $S=1/2$ electron spins~\cite{degraafDirectIdentificationDilute2017} at a density matching that of the surface spins producing the ubiquitous $1/f$ flux noise in SQUIDs~\cite{sendelbachMagnetismSQUIDsMillikelvin2008}. We expect saturation of noise suppression from spin freezing to occur in the regime $\frac{\gamma_e}{2 \pi} B \gg k_B T_\text{eff} / h \gtrsim \SI{800}{\mega Hz}$, where $\gamma_e/2\pi\approx\SI{2.8}{\mega Hz/G}$ is the free-electron gyromagnetic ratio and the lower bound of $T_{\text{eff}}$ is set by the measured mixing-chamber plate temperature in our experiment ($\approx\SI{40}{\milli K}$). Our largest applied field ($B_\text{max}=\SI{100}{G}$) corresponds to a free electron Zeeman energy of $\frac{\gamma_e}{2 \pi} B_\text{max} \approx \SI{280}{\mega Hz}$ which is below the thermal energy scale of $\approx \SI{800}{MHz}$. Given the non-monotonic behavior of the low-frequency flux noise and the saturation of the high-frequency spin-echo dephasing, we suggest that thermal polarization alone cannot explain the observed trends.

One plausible interpretation of the data is an increase in surface spin cluster size with magnetic field, where a cluster refers to a group of interacting spins. In this paragraph, we justify clustering as a relevant phenomenon in superconducting qubit surface spin baths.  Clustering is a known behavior of spin ensembles in proximity to a phase transition~\cite{atalayaFluxEnsuremathAlpha2014, deIsingGlauberSpinCluster2014}, with the cluster-size distribution depending on temperature~\cite{atalayaFluxEnsuremathAlpha2014} and field~\cite{diaz-mendezPhaseDiagramTwodimensional2010}. Multiple experiments on similar superconducting quantum circuits have observed evidence of native surface spin baths being near a magnetic phase transition while at standard operating conditions (i.e. millikelvin temperatures and nominally zero applied field)~\cite{sendelbachMagnetismSQUIDsMillikelvin2008, lantingProbingEnvironmentalSpin2020}.  
Increasing cluster size has been previously hypothesized as a source of the spectral pivoting of $1/f$ noise with decreasing temperature~\cite{antonMagneticFluxNoise2013}.
We note that experimental and theoretical studies have suggested that high-frequency $1/f$ flux noise emerges from spin diffusion dynamics~\cite{faoroMicroscopicOriginLowFrequency2008, atalayaFluxEnsuremathAlpha2014, lantingProbingEnvironmentalSpin2020,lantingEvidenceTemperaturedependentSpin2014}, while a distinct mechanism is responsible for the low-frequency flux noise, such as longer-time fluctuations of the net magnetization of clusters~\cite{atalayaFluxEnsuremathAlpha2014, lantingProbingEnvironmentalSpin2020}.

We now discuss the low frequency ($\lesssim \SI{10}{Hz}$) flux noise spectrum [Fig.~\ref{fig:fig3}a]. It has been suggested that clusters of spins may act as ``macrospins" with effective magnetic moments and relaxation processes, which produce an ensemble of telegraphic noise processes giving rise to $1/f$ noise~\cite{atalayaFluxEnsuremathAlpha2014, deIsingGlauberSpinCluster2014, deFluxNoiseLow2019}. Assuming the effective relaxation rate of a cluster rapidly decreases with the number of spins in the cluster~\cite{atalayaFluxEnsuremathAlpha2014}, an increasing size with applied field would be consistent with the rise in low-frequency flux noise. The transition from $1/f$ to approximately Lorentzian noise suggests a narrowing of the distribution of cluster relaxation rates, which may reflect clusters becoming more homogeneous in size with applied field as a result of, e.g., fewer total clusters or size saturation due to the finite dimension of the superconducting wire. 
We note that the Lorentzian cut-off frequency does not saturate with magnetic field, but appears highest at $B\approx\SI{80}{G}$. This non-monotonic behavior may be due to the field dependence of system parameters such as individual spin relaxation times, the effective spin diffusion constant, cluster sizes, etc. 

We now proceed to the high frequency ($\gtrsim \SI{1}{\mega Hz}$) flux noise spectrum [Fig.~\ref{fig:fig3}b]. 
We present two potential mechanisms for the suppression of spin diffusion that would lead to the lowering of flux noise in the MHz range: (1) spin clustering, and (2) inhomogeneous broadening of the spin bath. The growth of spin clusters (and corresponding reduction of their flip-rate~\cite{ atalayaFluxEnsuremathAlpha2014}) would reduce the number of smaller clusters contributing to high-frequency noise~\cite{antonMagneticFluxNoise2013}. Beyond this generic trend, we consider the case of ferromagnetic or random clusters. In the case of ferromagnetic clusters, growth would inhibit flip-flop processes contributing to spin diffusion by decreasing the number of participating antiparallel spin pairs. In the case of random clusters (in which spins are oriented randomly), growth would inhibit diffusion processes past a critical timescale $f_c^{-1} \propto L^2/D$, where $D$ is the effective spin diffusion coefficient and $L$ is the spatial extent of a cluster which determines how far excitations can freely diffuse before running into a boundary---at frequencies above $f_c$, the noise PSD asymptotically approaches $S(f>f_c) \sim 1/f^{1.5}$~\cite{lantingEvidenceTemperaturedependentSpin2014, lantingProbingEnvironmentalSpin2020}. 
In addition to a reduction of the noise level, our data displays an increasing noise exponent $\alpha$ with applied field which is consistent with an increase in $L$. We note that in an earlier cooldown we observed $\alpha\approx1.5$ at magnetic fields $B\gtrsim\SI{12}{G}$ in multiple datasets, although this behavior was not observed during the subsequent cooldown. We also note the apparent saturation of $\Gamma^{E}_{\phi}$, which may suggest a saturation of spin cluster sizes complementing the $1/f$ to Lorentzian transition in the low-frequency noise. 

Another possible mechanism for the suppression of spin diffusion (i.e. spin flip-flops) is inhomogeneous broadening of the spin bath from local variations in the applied field~\cite{ bermanSpinDiffusionRelaxation2005, tyryshkinElectronSpinCoherence2012b}, which would reduce the effective diffusion constant $D$~\cite{bloembergenInteractionNuclearSpins1949}. Spin flip-flops are possible between resonant spins (detuned less than their interaction strength) with antiparallel orientations. 
Since the saturation of $\Gamma^{E}_{\phi}$ occurs at lower field than would be expected from polarization (reorientation) of the surface spins, we suggest that inhomogeneous broadening provides a more consistent explanation for both the qubit coherence and noise spectroscopy data.
A number of mechanisms may lead to inhomogeneous broadening such as spatially inhomogeneous Meissner screening, or a statistical distribution of the effective gyromagnetic ratios of magnetic defects. Attributing the saturation of $\Gamma^{E}_{\phi}$ entirely to such broadening, we place a rough bound on the spin-spin interaction strength assuming the spin energy is given approximately by the applied field Zeeman splitting---a spin experiencing the bare field would have a frequency $\approx\frac{\gamma_e}{2\pi} B$, and a nearby spin experiencing no applied field (i.e. on an adjacent face of the wire which is entirely shielded) would have a frequency $\approx0$. Flip-flop processes would be inhibited between these spins if their coupling strength $J$ satisfies $J < h \frac{\gamma_e}{2\pi} (B - 0)$. With a saturation field $B_\text{sat}\approx\SI{50}{G}$, we have $J/h \lesssim 150$ MHz.

In summary, our results provide the first characterization of flux-noise-limited dephasing in superconducting qubits as a function of applied magnetic field. Our data reveals a distinct $1/f$ to approximately Lorentzian transition of the noise spectrum below \SI{10}{Hz} as well as a suppression of noise above \SI{1}{MHz}. The observed trends are consistent with increasing spin cluster sizes with applied field, although more experimental and theoretical investigation is required to validate this interpretation. Further insight can be obtained by mapping the flux noise response at higher fields using magnetic-field-resilient devices (e.g. niobium or thinner aluminum metallizations), or by probing the noise response to applied fields while varying device materials or field angle. In addition, searching for resonant peaks in the flux noise spectra at higher frequency ($\gtrsim\SI{10}{MHz}$) as a function of magnetic field may provide valuable clues about the electronic and chemical configuration of the magnetic defects comprising the spin bath. Such signatures of coherent fluctuators in flux noise spectra have already been observed, albeit at nominally zero field, and without consistent reproducibility~\cite{yanRotatingframeRelaxationNoise2013}. Already, we anticipate that our results can provide a new experimental constraint for future flux noise models incorporating magnetic field dependence, which may bring us one step closer to solving the decades-long open question of the microscopic origin of $1/f$ flux noise in superconducting circuits. 

We gratefully acknowledge Patrick Harrington, Agustin Di Paolo, Amir Karamlou, Youngkyu Sung, Antti Vepsalainen, Tim Menke, Ilan Rosen, James Ehrets II, Greg Calusine, Charlotte B\o ttcher, and Rogério de Sousa for many fruitful discussions and feedback. This material is based upon work supported by the U.S. National Science Foundation (NSF), the U.S. Department of Energy, the Under Secretary of Defense for Research and Engineering (under Air Force Contract No. FA8702-15-D-0001), the Office of Science National Quantum Information Science Research Center's Co-design Center for Quantum Advantage (contract no. DE-SC0012704). D.A.R. acknowledges support from the NSF (award DMR-1747426). D.A.R. and L.A. acknowledge support from the NSF Graduate Research Fellowship (grant no. 1745302). D.B. acknowledges support from the NSF Graduate Research Fellowship Program (grant DGE1745303) and The Fannie and John Hertz Foundation. The views and conclusions contained herein are those of the authors and should not be interpreted as necessarily representing the official policies or endorsements, either expressed or implied, of the U.S. Government.

\bibliography{refs}

\newpage
\onecolumngrid
\begin{center}
    \textbf{SUPPLEMENTAL MATERIAL}
\end{center}

\setcounter{figure}{0}
\setcounter{equation}{0}
\makeatletter 
\renewcommand{\thefigure}{S\@arabic\c@figure}
\renewcommand{\thetable}{S\@arabic\c@table}
\renewcommand{\theequation}{S\arabic{equation}}

\makeatother

\begin{section}{Device and Experimental Setup}
\begin{figure}[h!]
    \centering
    \includegraphics[width=\textwidth]{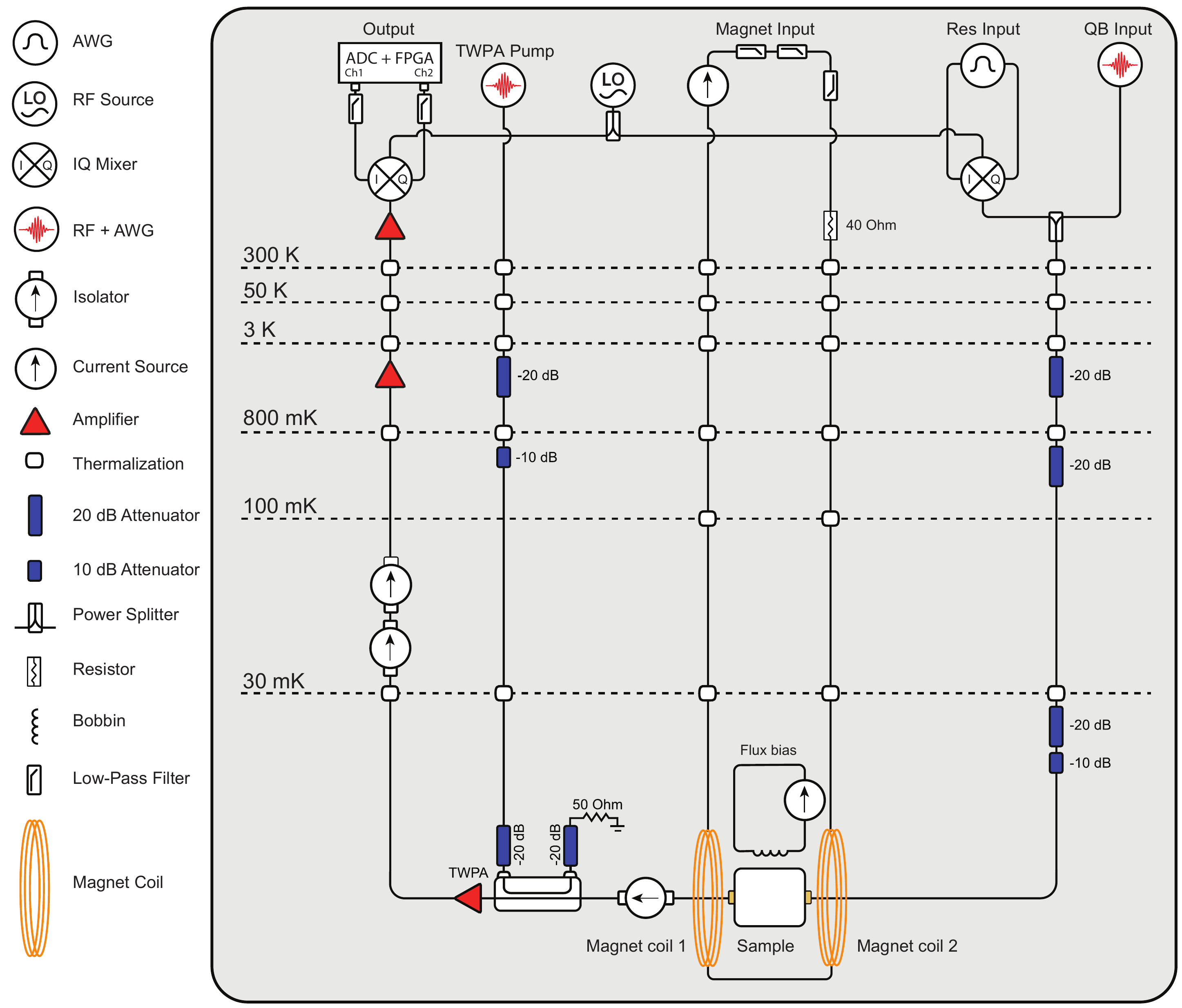}
    \caption{Experimental setup.}
    \label{fig:exp_setup}
\end{figure}

The experiment was conducted in a Leiden CF-650 dilution refrigerator (DR) operating at a base temperature of 30-\SI{40}{mK}. The sample was mounted on a cold finger with the magnet, both inside a superconducting Aluminum magnetic shield. A Mu-metal encasing around the base stage provided a second layer of magnetic shielding. The experimental setup is shown in Fig.~\ref{fig:exp_setup}. A common DC flux bias for tuning the qubit frequency was provided to the whole chip using a 500-turn bobbin attached to the package lid, biased by a QDevil QDAC. Input signals providing qubit control and readout pulses were generated by arbitrary waveform generators (AWGs, one Keysight M3202A for qubit signals and another for resonator signals) and RF sources (Rohde \& Schwarz SGS100 for the qubit, Agilent E8257C for the resonator), upconverted with IQ mixers, and sent into the DR. Output signals were downconverted and digitized with an ADC (Keysight M3102A). A Keysight M9019A chassis was used to synchronize and trigger the AWGs and ADC.

\begin{table}[h!]
    \centering
    \begin{tabular}{|c|c|c|}
        \hline
        Component           & Manufacturer     & Model   \\ \hline
        Dilution Fridge     & Leiden           & CF-650  \\
        RF Source (Qubit)   & Rohde \& Schwarz & SGS100  \\
        RF Source (Readout) & Agilent          & E8257C  \\
        DC Source (Qubit)   & QDevil           & QDAC    \\
        DC Source (Magnet)  & Yokogawa         & GS200   \\
        Control Chassis     & Keysight         & M9019A  \\
        AWG                 & Keysight         & M3202A  \\
        ADC                 & Keysight         & M3102A  \\ \hline
    \end{tabular}
    \caption{\textbf{Summary of control equipment.} The manufacturers and model numbers of the control equipment used for the experiment.}
    \label{tab:equipment}
\end{table}

\end{section}

\subsection{Sample}
The sample, originally characterized in \cite{braumullerCharacterizingOptimizingQubit2020}, comprises 10 uncoupled Al flux qubits with individual dispersively coupled resonators for control and readout, all multiplexed to a single transmission line. SQUID loop parameters were varied across the chip; the measured qubit had a rectangular SQUID with inner dimensions $\SI{18.32}{\micro m} \times \SI{90.51}{\micro m}$ and wire width of $\SI{1}{\micro m}$. The flux noise amplitude in the experimental setup at $B=0$ G was measured to be $A_\Phi \approx (\SI{4.95}{\micro \Phi_0})^2$ with the method of~\cite{braumullerCharacterizingOptimizingQubit2020}, within $10\%$ of the value measured in the previous experiment which was conducted in a different dilution refrigerator with different control electronics over one year ago at the time of submission. This supports the hypothesis of intrinsic flux noise limited dephasing for the device, rather than extrinsic flux noise limitations from the setup.

\subsection{Magnet}
Two NbTi superconducting coils in a Helmholtz geometry (hand-wound, 868 turns for each coil) provided in-plane fields of up to $B = \SI{100}{G}$ at an always-on bias current of \SI{540}{mA}, supplied by four Yokogawa GS200 DC sources in a parallel current-source configuration. The field per current at the location of the sample was analytically calculated from the geometry of the coils and verified with a Lakeshore F71 Teslameter at room temperature. Room-temperature filtering on the current bias lines ensured that noise coming from the power supply did not dominate the qubit flux noise (see Supplementary Material~\ref{supp:field_noise}). Due to field misalignment, a slight out-of-plane component $B_\perp$ provided an additional flux offset to the qubits. From measuring the periodicity of 5 separate qubit spectra on the same chip with respect to the applied field, we estimated the field misalignment to be $\epsilon \approx 0.2 \pm 0.05 \%$ (where $B_\perp = \epsilon B$). When sweeping the magnetic field, the qubit spectrum and flux noise susceptibility $\partial f_{01} / \partial \Phi$ was recalibrated with the separate global bobbin flux bias.

\begin{section}{Applied Magnetic Field}\label{supp:field}

\begin{subsection}{Evolution of Qubit Spectrum and Susceptibility to Applied Field Noise}
We first note the response of the qubit sweet spot frequency to the in-plane field. Placing a Josephson junction in a magnetic field results in the suppression of the critical current $I_c$, which follows a Fraunhofer pattern as a function of the applied field. This phenomena has been observed to affect both the intended junctions as well as the large-area parasitic junctions resulting from shadow evaporation processes in device fabrication~\cite{schneiderTransmonQubitMagnetic2019}. We observe this effect and present the measured sweetspot frequency as a function of field in Fig.~\ref{fig:figS2}a.

We then note that, at specific magnetic fields, there are divergences in the sweet spot frequency  indicating a heightened susceptibility of the qubit frequency to noise in the applied field, even at the sweet spot. We display the gradient computed for fit values of the sweet spot frequency as a function of field in Fig.~\ref{fig:figS2}b. We note that at the working point of $B=\SI{40}{G}$, the qubit sweet spot frequency is highly susceptible to noise in the applied field. This provides a likely explanation for the heightened dephasing noise observed at the sweet spot at $B=\SI{40}{G}$ in the data of Fig.~2. In addition, we probe the sensitivity of the qubit frequency to the applied field while moving off of the sweet spot in Fig.~\ref{fig:figS2}c, and observe the highest sensitivity at $B=\SI{40}{G}$. This heightened susceptibility to applied field provides a likely explanation for the outlier dephasing rates observed off the sweet spot at $B=\SI{40}{G}$ in Fig.~2. 

\begin{figure}[h!]
    \centering
    \includegraphics[width=0.8\textwidth]{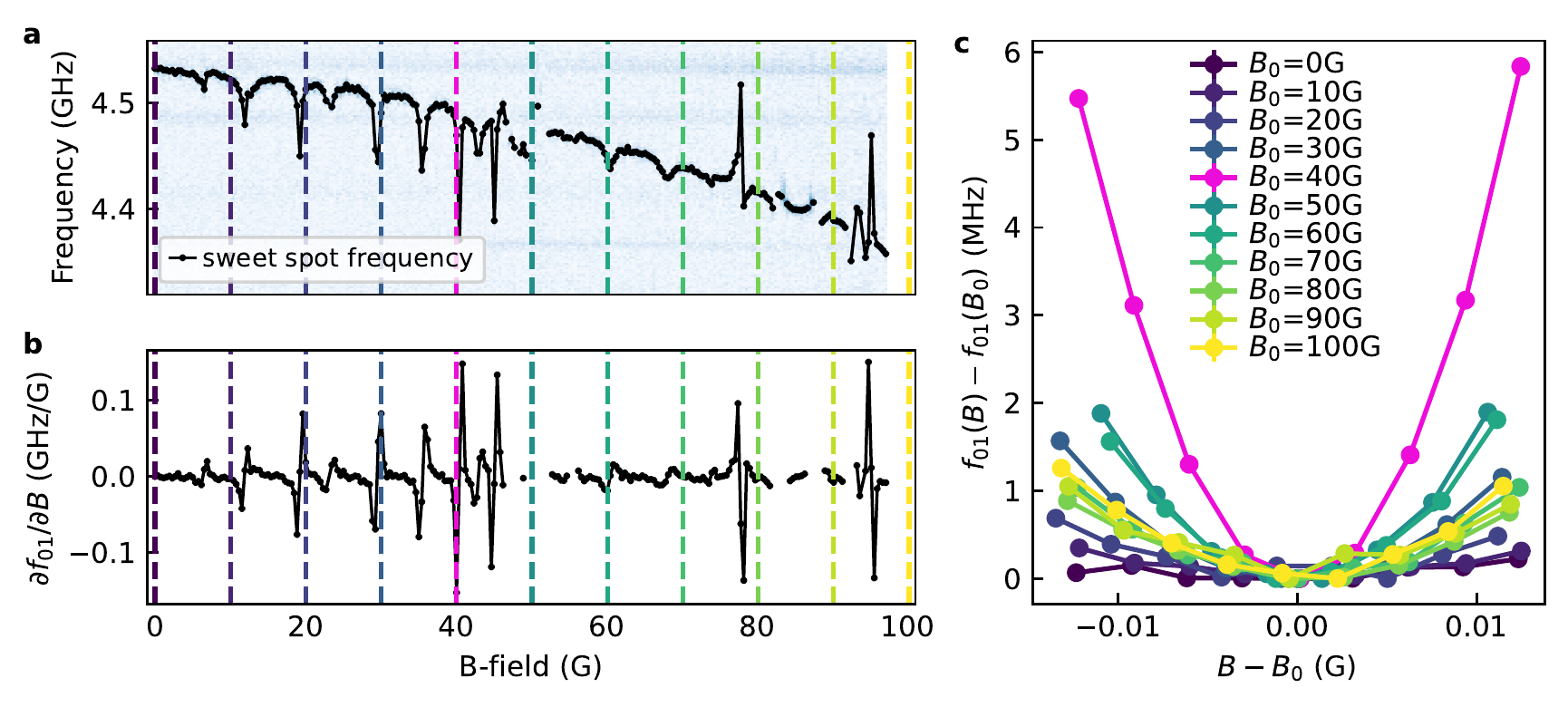}
    
    \caption{\textbf{Evolution of the Qubit Spectrum and Susceptibility to Field Noise.} \textbf{a)} Flux qubit sweet spot frequency vs in-plane field. At each magnetic field, the external flux is set to $\Phi = \SI{0.5}{\Phi_0}$. \textbf{b)} Gradient of the sweet spot frequency vs in-plane field, indicating the susceptibility of the qubit frequency to in-plane field fluctuations while operating at the sweet spot. \textbf{c)} Qubit spectroscopy around the sweet spot as a function of in-plane field. At each field $B_0$, the qubit is moved to the sweet spot $\Phi = \SI{0.5}{\Phi_0}$ with an additional flux control. The in-plane field is then swept in a small range around $B_0$. Each trace is artificially offset to $f_{01}(B_0) = 0$. The slope of the spectrum indicates the sensitivity of the qubit frequency $f_{01}(B)$ to in-plane field fluctuations, with the highest susceptibility at fields closest to $B_0=\SI{40}{G}$.}
    \label{fig:figS2}
\end{figure} 

\end{subsection}

\begin{subsection}{Applied Field Noise}\label{supp:field_noise}
In order to guarantee that observed trends in the qubit flux noise are not artifacts from changes in the applied magnetic field noise, we directly measured fluctuations of the magnet bias current $I$ at both zero field and the highest field, shown in Fig.~\ref{fig:figS3}. The current noise PSD $S_I(f)$ is converted to units of flux noise by
\begin{equation}
    S_{\Phi}(f) = \left(\frac{\partial{f_{01}}}{\partial{\Phi}}\right)^{-2} \left(\frac{\partial{f_{01}}}{\partial{I}}\right)^2 S_{I}(f),
\end{equation}
where we measure $\partial{f_{01}}/\partial{I}$ at identical points on the qubit spectrum $\partial{f_{01}}/\partial{\Phi} = \SI{25.0}{\giga Hz/\Phi_0}$ for low and high fields separately. At the frequencies probed by single-shot Ramsey noise spectroscopy ($\lesssim$ \SI{10}{Hz}), we observe the magnetic field noise is well below the measured qubit noise at both low and high fields. At higher frequencies ($\gtrsim$ \SI{10}{\kilo Hz}), the applied field noise decreases below the noise floor of the instrumentation. In this region relevant for spin-locking spectroscopy, we bound the noise with the transfer function $H(f)$ of in-line filters (measured by injecting a known magnitude white voltage noise $S^\text{in}_V(f)$ to one end of the filter chain, and measuring the output noise after the filters, $S^\text{out}_V(f) = H(f)\cdot S^\text{in}_V(f)$), which places the projected applied field noise $S^{\text{projected}}_\Phi(f) = H(f) \cdot S^{\text{unfiltered}}_\Phi(f)$ several orders of magnitude below the measured qubit flux noise at $B=\SI{0}{G}$, where $S^{\text{unfiltered}}_\Phi(f)$ corresponds to the measured field noise without the in-line filters. We were unable to measure the transfer function of the in-line filters while applying the bias current required for $B=\SI{100}{G}$, but we observe that the current noise measurement floor corresponds to $\lesssim7\%$ of the lowest measured qubit noise at all fields ($S_\Phi(\SI{8.3}{MHz}) \approx \SI{4.7e-17}{\Phi_0^2/Hz}$ at $B=\SI{20}{G}$). We note that in the configuration used to measure qubit flux noise, the total transfer function including the wiring of the fridge and inductance of the magnet would necessarily strengthen the bound given by the instrumentation floor, as there are only passive components in the magnet wiring. We also note that for all frequencies, the measured applied field noise increases with field, which is opposite to the observed qubit flux noise decreasing with field in the frequency domain relevant for spin-locking spectroscopy and spin echo pure dephasing. This further supports the independence of the observed qubit flux noise trends from the applied field noise.

\begin{figure}[ht]
    \centering
    \includegraphics[width=0.8\textwidth]{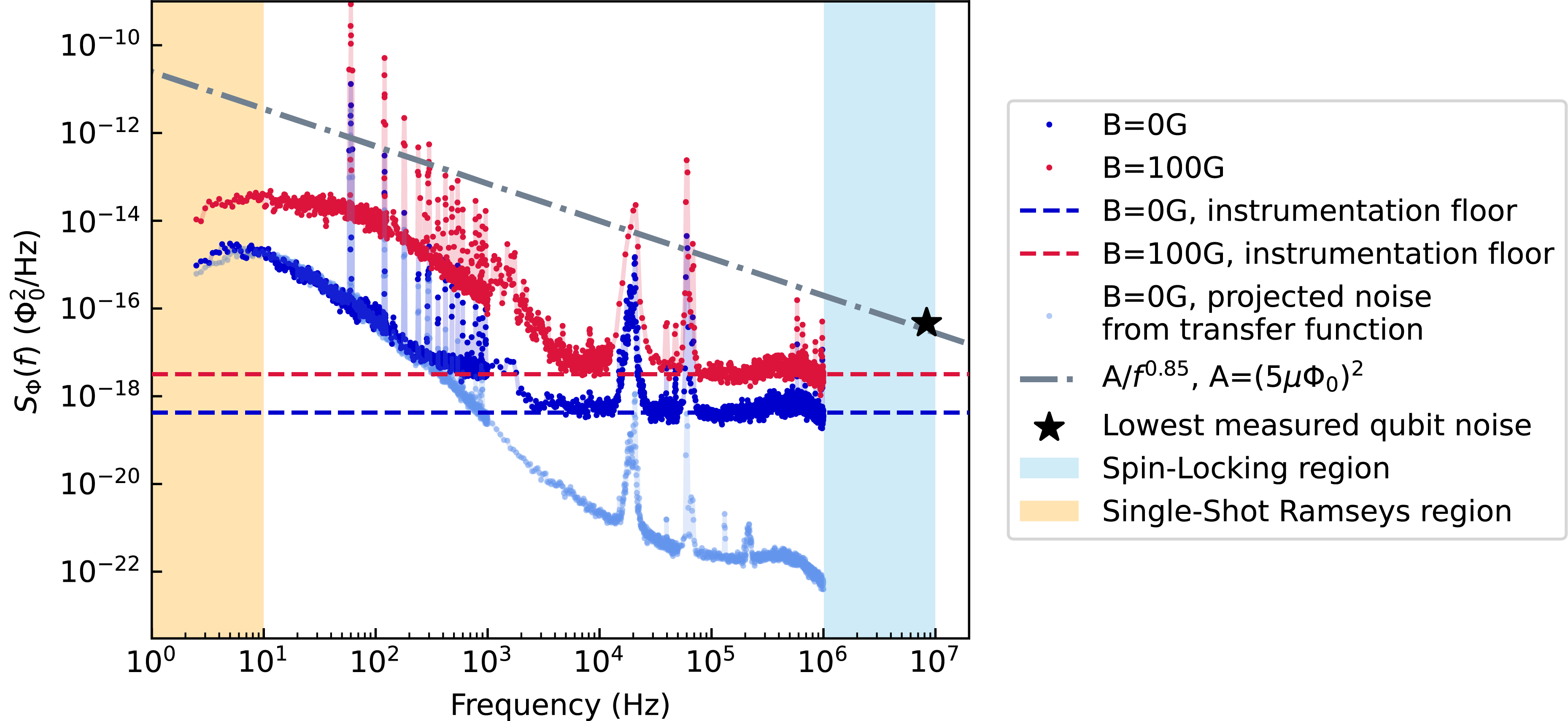}
    
    \caption{\textbf{Applied magnetic field noise.} Measured field noise at the lowest ($B=\SI{0}{G}$) and highest ($B=\SI{100}{G}$) fields is shown. The instrumentation noise floor is shown for both fields (dotted horizontal lines). We measure the transfer function of the in-line filters with zero bias current (corresponding to $B=\SI{0}{G}$), and use this to project the applied field noise below the instrumentation noise floor (light blue). We show an estimate of the qubit flux noise (gray dash-dotted line), as well as the lowest measured flux noise ($S_\Phi(\SI{8.3}{MHz}) \approx \SI{4.7e-17}{\Phi_0^2/Hz}$ at $B=\SI{20}{G}$, black star). Colored regions on the plot correspond to the frequency ranges of measured noise spectra via single-shot Ramseys (light orange) and spin-locking (light blue).}
    \label{fig:figS3}
\end{figure} 

\end{subsection}
\end{section}

\begin{section}{Noise Spectroscopy Methods}

\begin{subsection}{Pure Dephasing Rates}\label{supp:dephasingrates}
At each working point, the qubit lifetime $T_1=1/\Gamma_1$ is measured by preparing the qubit in the excited state and fitting the decay with an exponential function $p(t) = A\exp(-\Gamma_1 t) + C$. The Ramsey and spin-echo coherences are then measured to extract the pure dephasing rates $\Gamma_\phi^{R (E)}$. At the sweet spot ($\partial f_{01} / \partial \Phi = 0$), we observe exponential decays of the Ramsey (spin-echo) coherence for all magnetic fields and extract the pure dephasing rate by fitting the decay to $p(t) = A\exp[-(\Gamma^{R(E)}_\phi + \Gamma_1/2) t]\cdot f_{R(E)}(t) + C$ where $\Gamma_1$ is a fixed parameter determined by the preceding relaxation measurement, $f_E(t)=1$ and $f_R(t) = \sin(2\pi f t + \delta)$, where $f$ is an oscillation frequency equal to the detuning of the qubit and the applied pulse frequencies.

At the operating point $\partial f_{01} / \partial \Phi = \SI{26.0}{\giga Hz/\Phi_0}$, we assume $1/f$ noise limited dephasing which leads to Gaussian decay envelopes for both Ramsey and spin-echo experiments. For Ramsey experiments at all fields, we detune our qubit drive to create oscillations for ease of fitting the decay envelope. Surprisingly, in the range of magnetic fields from $B\approx\SI{50}{G}$ to $B \approx \SI{90}{G}$ we observe beating in the oscillations which are best fit with three oscillatory components. Some representative Ramsey decay traces are displayed in Fig.~\ref{fig:figS4}. In particular, we extract the pure dephasing rate $\Gamma^{R(E)}_\phi$ by fitting the Ramsey (spin-echo) decay to $p(t) = A \exp[-\Gamma_1 t / 2 - (\Gamma^{R(E)}_\phi t)^2] \cdot f_{R(E)}(t) + C$ where for spin-echo traces $f_E(t)=1$ for all fields and for Ramsey traces $f_R(t) = \sin(2\pi f t + \delta)$ for fields outside of the interval $B=\SI{50}{G}$ to $B=\SI{90}{G}$, and $f_R(t) = \sin(2\pi f_1 t + \delta_1) + \sin(2\pi f_2 t + \delta_2) + \sin(2\pi f_3 t + \delta_3)$ for $\SI{50}{G} \leq B \leq \SI{90}{G}$. For all fits, $\Gamma_1$ is a fixed parameter determined by a preceding $T_1$ measurement.

Due to calibration drifts and jumps in the middle of magnetic field sweeps, a criteria for rejecting data collected with faulty calibration was used. The criteria was independent of magnetic field, and consisted of checking that the data fit with the appropriate fit function yielded a coefficient of determination $R^2 > 0.9$.

\begin{figure}[ht]
    \centering
    \includegraphics[width=0.8\textwidth]{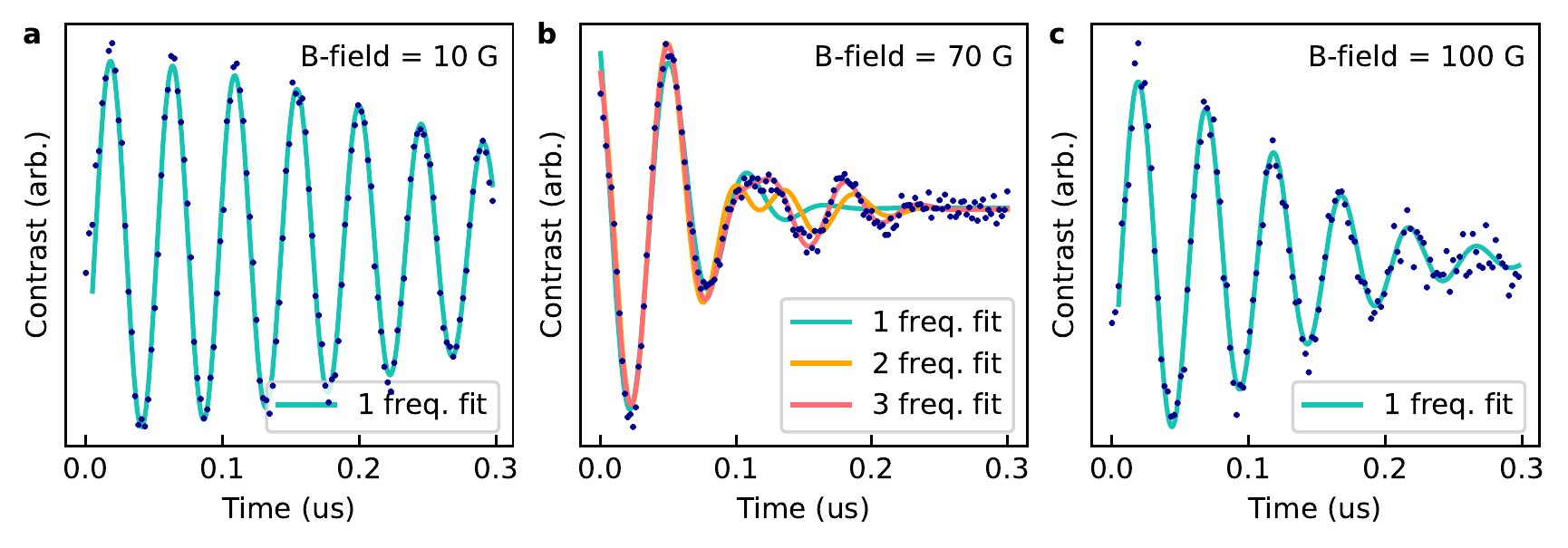}
    
    \caption{\textbf{Ramsey beating in a magnetic field.} Representative Ramsey decay traces at \textbf{a)} $B=\SI{10}{G}$, \textbf{b)} $B=\SI{70}{G}$, and \textbf{c)} $B=\SI{100}{G}$. All data is truncated at a time of \SI{0.3}{\micro s}. The Ramsey decays for magnetic fields from $B=\SI{50}{G}$ to $B=\SI{90}{G}$ displayed beating and were best fit by decay functions with three oscillatory components, detailed in Supplemental Material~\ref{supp:dephasingrates}.}
    \label{fig:figS4}
\end{figure} 

\end{subsection}

\begin{subsection}{Low-Frequency ($\lesssim\SI{10}{Hz}$) Noise Spectroscopy}

For low-frequency noise spectroscopy, we measured $S_{f_{01}}(f)$ with the single-shot Ramsey method detailed in~\cite{yanSpectroscopyLowfrequencyNoise2012} which utilizes a time series of repeated single-shot qubit-state measurements following a fixed-time free induction decay in the presence of dephasing noise. The single-shot readout voltages were classified with a state discriminator to obtain a binary time series and converted to flux noise by a calibrated scale factor, utilizing a cross-correlation calculation to remove statistical sampling noise~\cite{yanSpectroscopyLowfrequencyNoise2012}. 

We attribute the white noise floor to readout infidelity, modeled by uncorrelated errors in the qubit state readout with a probability $(1 - F) / 2$, yielding a voltage white noise floor $S_V^{\text{ro}}(f) = (1-F^2) \Delta t \cdot d^2$~\cite{risteMillisecondChargeparityFluctuations2013}, where $\Delta t$ is the repetition time of single-shot measurements, and $d$ is the voltage separation of the ground and excited state cluster means. For all measured PSDs, the measured noise floor yields $F \gtrsim 85\%$ which is consistent to within $20\%$ of the separation fidelity $F$ estimated by a 2-state Gaussian mixture model applied to the single-shot data for all datasets.

All presented PSDs are comprised of an average of 10 sequentially measured PSDs and smoothed with a rolling window of 7 points in the frequency domain. To measure $S_\Phi(f)$, we choose an operating bias slope of \SI{22.0}{\giga Hz/\Phi_0} (\SI{21.0}{\giga Hz/\Phi_0} at $B=\SI{100}{G}$) to obtain high sensitivity to frequency noise while maintaining reasonable readout SNR at high fields. To confirm that the measured noise spectra off the sweet spot and their magnetic field dependence are dominated by flux noise, we measure $S_{f_{01}}(f)$ at the sweet spot and show that it is both field-independent and well below the PSD at the flux-sensitive point for all magnetic fields in the frequency range $f \lesssim \SI{10}{Hz}$ (Fig.~\ref{fig:figS5}a).

Fit functions and resulting parameters for the data of Fig.~3a are presented in Table~\ref{tab:fig3fits}.
\begin{table}[h!]
    \centering
    \begin{tabular}{|l|l|l|}
        \hline
        Fit Function            & $B$ (G)  & Fit Parameters   \\ \hline
        $S_\Phi(f) = A/f^\alpha + C$                    &  0  & $A=\SI{1.65\pm.002}{\micro\Phi_0^2}$, $\alpha=\SI{0.987\pm0.004}{}$, $C=\SI{1.205\pm.006}{\micro\Phi_0^2 / \Hz}$    \\ \hline
                                                        & 20  & $\Gamma=\SI{0.438\pm0.003}{1/s}$, $A=\SI{97.1\pm0.5}{\micro\Phi_0^2/\Hz}$, $C=\SI{1.41\pm0.02}{\micro\Phi_0^2/\Hz}$ \\ 
        $S_\Phi(f) = \frac{A}{(1 + (2f/\Gamma)^2)} + C$ & 50  & $\Gamma=\SI{0.890\pm0.006}{1/s}$, $A=\SI{239\pm1}{\micro\Phi_0^2/\Hz}$, $C=\SI{2.24\pm0.07}{\micro\Phi_0^2/\Hz}$    \\ 
                                                        & 80  & $\Gamma=\SI{1.301\pm0.009}{1/s}$, $A=\SI{309\pm2}{\micro\Phi_0^2/\Hz}$, $C=\SI{1.8\pm0.1}{\micro\Phi_0^2/\Hz}$      \\ 
                                                        & 100 & $\Gamma=\SI{0.631\pm0.003}{1/s}$, $A=\SI{219.5\pm0.9}{\micro\Phi_0^2/\Hz}$, $C=\SI{1.78\pm0.03}{\micro\Phi_0^2/\Hz}$      \\ \hline
        
    \end{tabular}
    \caption{\textbf{Summary of fit functions and parameters for Fig.~3a .}}
    \label{tab:fig3fits}
\end{table}

\end{subsection}

\begin{subsection}{High-Frequency ($\gtrsim\SI{1}{MHz}$) Noise Spectroscopy}

For high-frequency noise spectroscopy, we use the spin-locking method detailed in~\cite{yanRotatingframeRelaxationNoise2013}, with interleaved SL-5a, SL-5b, and $T_1$ pulses. In order to average away time-dependent fluctuations in the noise environment, we avoid taking frequency data sequentially, instead taking frequency points in 10 interleaved sequences. To obtain the PSD of the qubit pure dephasing noise at a frequency $f$, we apply the SL-5a and SL-5b pulse sequences with Rabi frequency $f_R = f$, and fit the resulting averaged trace to an exponential decay with rate $\Gamma_{1\rho} = \Gamma_{\nu} + \Gamma_1/2$, where we obtain $\Gamma_1$ from the interleaved $T_1$ sequence. The resulting decay rate $\Gamma_\nu$ is related to qubit frequency noise PSD (corresponding to noise along the undriven qubit quantization axis) by $\Gamma_\nu = 1/2 \cdot S_{f_{01}} (f=f_R)$. At the sweet spot, the frequency noise is well below that measured off the sweet spot, shows no distinct magnetic field dependence, and is approximately $T_1$ limited ($T_{1\rho} \gtrsim T_1$). When away from the sweet spot where the dephasing is dominated by flux noise, we plot data with the conversion $S_{f_{01}}(f) = (\partial f_{01} / \partial \Phi)^{2} \cdot S_\Phi(f)$.

\end{subsection}

\begin{figure}[ht]
    \centering
    \includegraphics[width=0.8\textwidth]{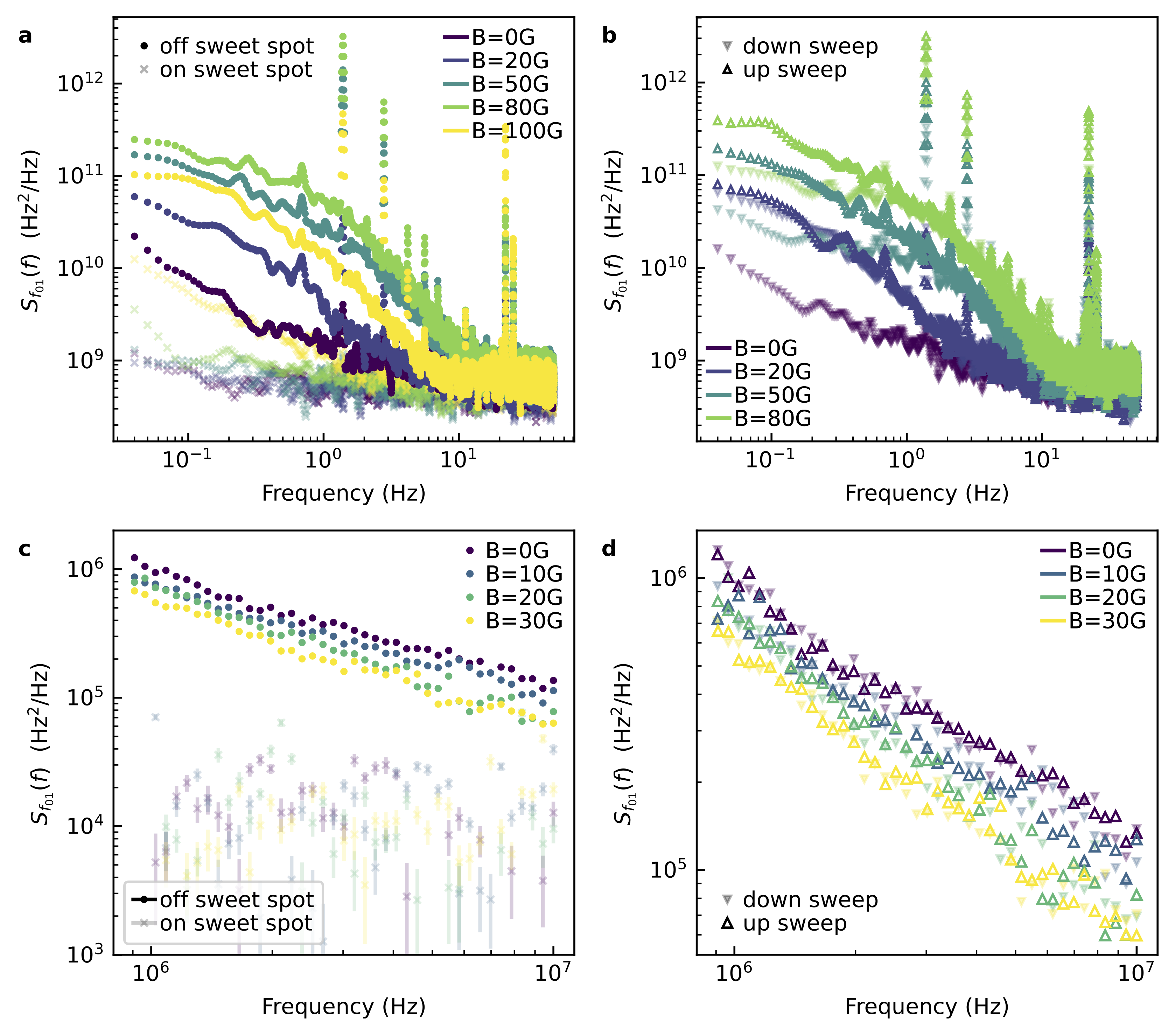}
    
    \caption{\textbf{Noise spectroscopy at the sweet spot and hysteretic effects.} PSDs are presented here in units of frequency noise, and for off sweet spot data are related to flux noise spectra by $S_{f_{01}}(f) = (\partial f_{01} / \partial \Phi)^2 S_{\Phi}(f)$. Off sweet spot data in \textbf{a,c} are the same as those presented in Fig.~3, and sweet spot data were collected in separate upwards field sweeps. \textbf{a)} Low frequency noise spectroscopy on and off the sweet spot as a function of magnetic field. \textbf{b)} Hysteresis of the low frequency noise spectra. All data was taken with $\partial f_{01} / \partial \Phi = \SI{22.0}{\giga Hz/\Phi_0}$. The field was first lowered from $B=\SI{80}{G}$ to $B=\SI{0}{G}$ (partially transparent downwards triangle markers), then raised back up to $B=\SI{80}{G}$ (faceless triangle markers). \textbf{c)} High frequency noise spectroscopy on and off the sweet spot as a function of magnetic field. \textbf{d)} Hysteresis of the high frequency noise spectra. Data was taken with $\partial f_{01} / \partial \Phi = \SI{30.0}{\giga Hz/\Phi_0}$ for $B\lesssim\SI{10}{G}$ and $\partial f_{01} / \partial \Phi = \SI{31.0}{\giga Hz/\Phi_0}$ for $B\gtrsim\SI{20}{G}$. The field was first raised from $B=\SI{0}{G}$ to $B=\SI{30}{G}$ (faceless triangle markers), then lowered back down to $B=\SI{0}{G}$ (partially transparent downwards triangle markers).}
    \label{fig:figS5}
\end{figure} 

\end{section}

\end{document}